\documentclass[11pt,twoside]{article}
\usepackage{asp2004}
\usepackage{epsfig} 

\begin{document} 

\title{Observing Dark Energy with SNAP} 

\author{Eric V.~Linder for the SNAP Collaboration} 
\affil{Physics Division, Berkeley Lab, 1 Cyclotron Road, 
Berkeley, CA 94720, USA} 

\begin{abstract} 
The nature of dark energy is of such fundamental importance -- 
yet such a mystery -- that a dedicated dark energy experiment should 
be as comprehensive and powerfully incisive as possible.  The 
Supernova/Acceleration Probe robustly controls for a wide variety of 
systematic uncertainties, employing the Type Ia supernova distance 
method, with high signal to noise light curves and spectra over 
the full redshift range from $z=0.1-1.7$, and the weak gravitational 
lensing method with an accurate and stable point spread function.  
\end{abstract} 

\section{Introduction} 

Exploring the nature of the physics responsible for the current 
acceleration of the expansion of the universe is a major scientific 
goal for the next decade.  The implications range over high energy 
physics, the theory of gravitation, and the fate of the universe. 
Current data, most notably Type Ia supernova distances, accompanied 
by cosmic microwave background (CMB) and large scale structure 
measurements, suggest a dark energy component comprising nearly 
three quarters of the total energy density, with negative pressure 
roughly equal to its energy density (i.e.\ pressure to energy 
density ratio $w\approx-1$). 

But the physics possibilities behind this are manifold.  Einstein's 
cosmological constant would have $w$ identically $-1$ at all redshifts 
$z$, while models exist with $w(z)$ greater than or less than $-1$ 
and generically time varying.  Moreover, many models possess an 
averaged value of $w$ not too different from $-1$ over a limited 
redshift range, so even a 5\% measurement of $\langle w\rangle$ or 
an assumed $w_{\rm constant}$ is unlikely to provide us with an 
important and unambiguous clue to fundamental physics. 

To progress, we need a new, specially designed experiment that seeks 
to account for and control all reasonably possible sources of uncertainty 
in the astrophysics, while probing the cosmology through multiple, well 
understood methods.  Ten years from now, we want to be in the position 
of having, if not the solution, then clear clues to solve the mystery 
of dark energy, but not to fail to obtain guidance after all that 
effort through lack of 
foresight and completeness.  The requirements of depth and sufficiency 
drive the design of the Supernova/Acceleration Probe \citep[SNAP;][]{snap}, 
a space based 2-meter 
telescope with optical and near infrared imaging and spectroscopy. 

\section{Beyond the Present} 

Ongoing ground based supernova surveys, necessarily limited to $z<1$, 
will improve our knowledge of 
$w_{\rm constant}$, and of supernovae themselves -- both extremely 
useful -- but do not give us directly the equation of state ratio of 
the dark energy.  From $w_{\rm constant}$ we cannot extract the present 
value $w_0$ or a measure of its variation $w'$ in an unbiased manner, and 
merely admitting our ignorance of whether there exists a variation $w'$ 
blows up the uncertainties in $w_0$ \citep{briansaul}. 

Space based surveys permit more accurate supernova photometry and the 
ability to extend to higher redshifts \citep[e.g.][]{knop03,riess04}, 
but the time is not available to follow up and characterize substantial 
numbers of supernovae (even with a dedicated program on the Hubble 
Space Telescope, due to its narrow field). 
So these advances, while 
solid steps of progress, are insufficient to obtain significant 
constraints on the dynamics of the dark energy through $w'$.  This lack 
of a dense, superbly characterized sample over the entire key redshift 
range of $z=0-1.7$ obscures and disables discrimination of the physics. 

Figure 1 illustrates this, showing the effect of a random irreducible 
systematic uncertainty of $0.1z$ per bin of redshift width 0.1.  This 
could arise from a combination of extinction and K-corrections, 
gravitational lensing magnification, calibration errors, etc.  Only 
an experiment obtaining scores of supernovae per bin over the full 
range, with exquisite systematics control, could achieve the 
sensitivity and power of the innermost, light yellow contour 
representing SNAP constraints at 68\% confidence level. 

\begin{figure}[!hbt]
\begin{center} 
\psfig{file=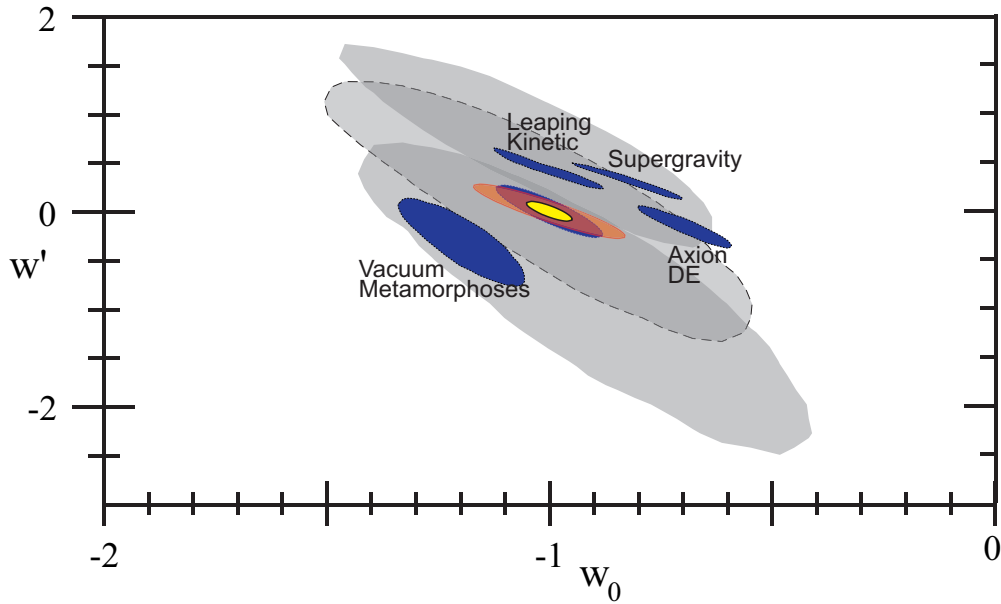,width=5.2in} 
\caption{Current and imminent supernova surveys do not have the redshift 
range, statistics, and systematic control necessary to discriminate 
even roughly between different models of dynamical dark energy.  The 
large central, dashed grey shaded region shows the current status, including 
HST supernovae, shifted to be centered on the cosmological constant 
model parameters ($\Omega_m=0.3\pm0.04$ prior included). 
The flanking grey shaded regions show the effects of bias induced by 
the current level of systematics, completely obscuring various dark 
energy models.  SNAP, as a new generation experiment, will be able to 
achieve the inner, light yellow contour. 
} 
\label{fig.grey5model}
\end{center} 
\end{figure}

Specifically, the requirements on the data sample are: 

\begin{itemize} 
\item Full redshift range $z=0-1.7$ with dense sampling, to 
break parameter degeneracies and bound systematics. 

\item $\sim$2000 supernovae with optical/near infrared imaging 
and spectra to 1) divide into subsets for like-to-like comparison 
(``anti-evolution''), 2) obtain high signal to noise to bound systematics 
and prevent Malmquist 
bias, and 3) many $z>1$ supernovae to prevent gravitational lensing bias. 

\item Space telescope ($\sim$2 meter aperture) for 1) infrared 
observations (essential for high $z$) and high accuracy color (dust 
extinction) corrections, and 2) precise and stable weak gravitational 
lensing shear measurements. 

\item Crosschecking and complementary methods for robust 
characterization of the nature of dark energy.  Weak lensing adds 
great value in deep, medium wide, and panoramic surveys.  {\it No need 
for $\Omega_M$ prior!} 

\end{itemize} 

\begin{figure}[!hbt]
\begin{center}
\psfig{file=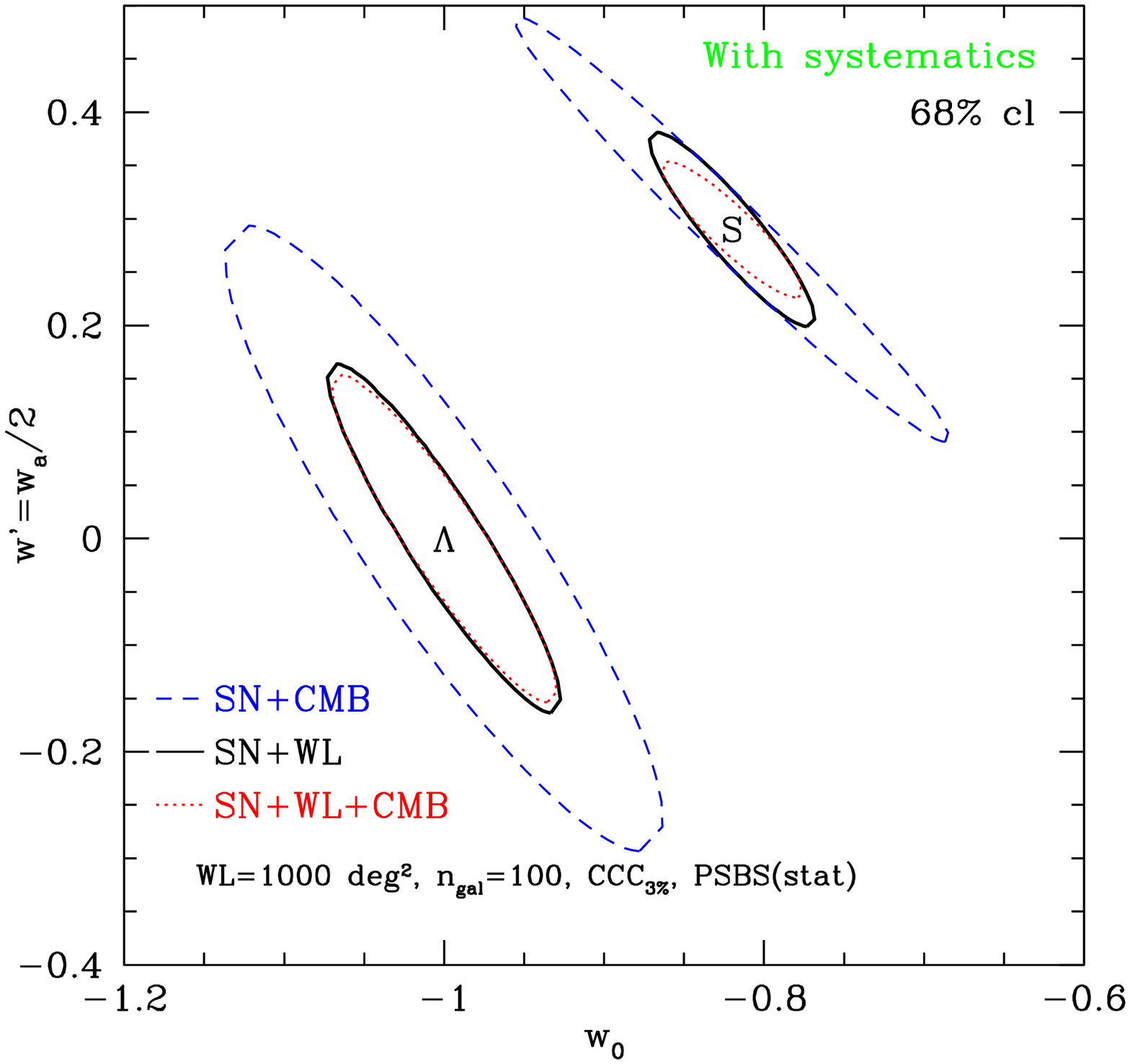,width=3.3in}
\caption{Weak gravitational lensing and supernovae distances work 
superbly together as cosmological probes.  To realize the tightest 
bounds requires systematics control only possible from space -- 
point spread function resolution, stability, and low noise.  Here 
we include constraints for two dark energy models 
from 2000 supernovae and a 1000 square degree 
weak lensing survey (employing power spectrum and bispectrum data 
and cross-correlation cosmography), both with systematics.  No 
external priors are needed. 
}
\label{fig.sncwall}
\end{center}
\end{figure}

\section{SNAP Surveys} 

SNAP plans its observing strategy to maximize the science from both 
the supernova and weak lensing methods.  In the reference mission, the 
deep survey covers 15 square degrees repeatedly in 9 wavelength bands 
for 120 visits, discovering and following supernovae down to AB 
magnitude 30.3 in each filter and measuring lensing shears for $10^7$ 
galaxies with a number density of greater than 250 resolved galaxies 
per square arcminute.  This will be superb for a wide area dark matter 
map.  The wide survey scans 300-1000 square degrees 
once, down to AB 26.6 in each band, resolving 100 galaxies per square 
arcminute for a total of some 300 million galaxies.  A panoramic 
survey is also under consideration of up to 10000 square degrees, AB=25, 
with 40-50 galaxies per arcmin$^2$ for a billion galaxies. 

\section{Observing Dark Energy} 

In order to determine accurately the cosmological parameters, the 
dark energy density $\Omega_{DE}$ and present equation of state $w_0$, 
and explore the crucial physics clue of the variation $w'$, SNAP will 
achieve: 

\quad$\bullet$ Homogeneously calibrated supernova sample over the 
full redshift range $z=0.1-1.7$; 

\quad$\bullet$ Tight systematics control; 

\quad$\bullet$ Unified, comprehensive mission without the need for 
external priors -- SN Ia + Weak Lensing (+ Strong Lensing + SN II +\dots); 

\quad$\bullet$ Precision {\it and} accuracy beyond proposed ground and 
non-dedicated space observations. 

Moreover, SNAP will test the cosmological framework by directly mapping 
the expansion history $a(t)$ of cosmic scales over time, from the acceleration 
into the deceleration epoch, probing dark energy, higher dimensions, 
extensions to gravity, etc. 

In addition, the science resources provided by the deep, wide, and 
panoramic fields will fuel a plethora of astronomical investigations 
into galaxy evolution, the high redshift universe, dark matter, stellar 
populations, rare, variable, and moving objects, etc.  Furthermore SNAP 
will act in synergy with the next generation James Webb Space Telescope 
and carry out a Guest Survey program. 

To observe dark energy and optimize the return on the investment of effort 
and funding by having the best chance for learning the essential dynamics 
to push the fundamental physics frontier, we need to have a 
comprehensive, robust experiment like SNAP. 

\acknowledgments{This work was supported in part by the Director, 
Office of Science, Department of Energy under grant DE-AC03-76SF00098. 
Special thanks to Daniel Holz for contributions to Figure 1 and to 
Masahiro Takada for contributions to Figure 2.}

\end{document}